\documentclass{article}

\usepackage{amsmath}

\usepackage{amssymb}

\usepackage{bm}
\usepackage{hyperref}

\begin{document}

\title{The existence of Bogomolny decomposition for baby Skyrme models}

\author{{\L}. T. St\c{e}pie\'{n} \thanks{The Pedagogical University of Cracow, ul. Podchora\c{}\.{z}ych 2, 30-084 Krak\'{o}w, Poland} \thanks{e-mail: sfstepie@cyf-kr.edu.pl, stepien50@poczta.onet.pl}}
             
        \date{} 
             
             \maketitle

\begin{abstract}
We derive the Bogomolny decompositions (Bogomolny equations) for: full baby Skyrme model and for its restricted version (so called, pure baby Skyrme model), in (2+0) dimensions, by using so called, concept of strong necessary conditions. It turns out that Bogomolny decomposition can be derived for restricted baby Skyrme model for arbitrary form of the potential term, while for full baby Skyrme model, such derivation is possible only for some class of the potentials. 
\end{abstract}



\section{Introduction}

The baby Skyrme model appeared firstly as an analogical model (on plane) to the Skyrme model in three-dimensional space. Since the target space of Skyrme model is $SU(2)$, \cite{Skyrme1961}, \cite{Skyrme1962}, \cite{Skyrme1971}, then for baby Skyrme model the target space is $S^{2}$. In these both models static field configurations can be classified topologically by their winding numbers.
Analogically to the Skyrme model, the baby Skyrme model includes: the quadratic term i.e. the term of nonlinear $O(3)$ sigma model, the  quartic term - analogue of the Skyrme term and the potential. The presence of the potential, in baby Skyrme model, is necessary, for existence of  static solutions with finite energy. However, the form of this potential is not restricted and different form of the potential were investigated in \cite{Leesetal1990},\cite{Pietetal1}, \cite{Pietetal2},\cite{Pietetal3},\cite{Sutcliffe1991}, \cite{Weidig1999}, \cite{Eslamietal2000}, \cite{Karlineretal2008}, some recent results are, among others, in \cite{Adametal2009}, \cite{Adametal2010}, \cite{JMSpeight2010}, \cite{Jaykkaetal2010}, \cite{Jaykkaetal2011}. In \cite{Ioannidouetal2009} noncommutative baby Skyrmions were studied.\\ 
The lagrangian of baby Skyrme model fas the form, \cite{Adametal2009}

 \begin{equation}
 \mathcal{L}=\partial_{\mu} \vec{S} \cdot \partial^{\mu} \vec{S} - \beta( \partial^{\mu} \vec{S} \times
 \partial^{\nu} \vec{S})^{2} - V(\vec{S}), \label{baby_Skyrme}
 \end{equation}
 
 where $\vec{S}$ is three-component vector field, such that $\mid \vec{S} \mid^{2} = 1$ and $\beta > 0$ is a coupling constant. 
The baby Skyrme model has simpler structure, than three-dimensional Skyrme model and so it can give an opportunity of better understanding of the solutions of Skyrme model in (3+1)-dimensions. Moreover, independently on it, the baby Skyrme model can be applied for the description of the quantum Hall effect, \cite{BelavinPolyakov1975}, \cite{Sondhietal2002}, \cite{Walet2001}. However, on the other hand, it is still complicated, non-integrable, topologically non-trivial and nonlinear field theory. Because of this reason, it is dificult to make analytical studies of this model and so, the investigations of baby Skyrmions have  very often numerical character. Therefore, the simplification, but of course, keeping us in the class of Skyrme-like models and simultaneosuly, giving an opportunity for analytical calculations, is important. One may, for example, try to define, which features of the solutions of the baby Skyrme model, are determined by which part of the model. So, one could neglect some particular part of the Lagrangian and so, investigate such simplified model. One may also simplify the problem of solving of field equations, by deriving Bogomolny equations (sometimes called as Bogomol'nyi equations) for these models, mentioned above. 
All solutions of Bogomolny equations satisfy Euler-Lagrange equations, which order is bigger than the order of Bogomolny equations. \\

In this paper we derive Bogomolny equations (we call them as Bogomolny decomposition) for these both models: restricted baby Skyrme and full baby Skyrme, in (2+0)-dimensions. This first one is characterized by absence of $O(3)$ term in (\ref{baby_Skyrme}).\\
The Bogomolny equations for restricted baby Skyrme model in (2+0)-dimensions, but for some special class of the potentials, was derived in \cite{Adametal2010}, by using the technique, firstly applied by Bogomolny in \cite{Bogomolny1976}, among others, for the nonabelian gauge  theory. This method is based on proper separation of the terms in the functional of energy. The solutions of Bogomolny equations, found in this way, minimalize the energy functional and saturate Bogomolny bound i.e. an inequality connecting energy functional and topological charge. 
In \cite{GP1997} the so called restricted (or pure) baby Skyrme model was studied, by Gisiger and Paranjape, which derived Bogomolny equations for the case, when the potential is $V(\vec{S})=(\vec{n} - \vec{S})^{2}=2(1-\vec{n}\cdot\vec{S})$, where $\mid \vec{n} \mid=1$ and $\vec{n}$ is a constant vector, selecting the vacuum. In \cite{Innocentisetal2001}, a second Bogomolny bound, for the model investigated previously in \cite{GP1997}, was found, as a contribution to some improved Bogomolny bound for the full baby Skyrme model. 

The Bogomolny bound for restricted baby Skyrme model in (2+0)-dimensions, derived in \cite{Adametal2010}, has the form

  \begin{gather}
   E = \frac{1}{2}\int d^{2}x \bigg( \bigg(\frac{1}{2} \epsilon_{ij}  \vec{S} \cdot (\partial_{i} \vec{S} \times
 \partial_{j} \vec{S}) \pm \gamma^{2} \sqrt{V(S^{3})} \bigg)^{2} \mp \label{bogomolny_Adam_etal:2010zz} \\ 
  \gamma \sqrt{V(S^{3})} \epsilon_{ij} \vec{S} \cdot ( \partial_{i} \vec{S} \times \partial_{j} \vec{S}) \bigg) \geq 4 \pi 
  \gamma C_{1} \mid Q \mid,  \nonumber 
  \end{gather} 
  
  where 
  
  \begin{equation}
  Q = \frac{1}{8 \pi } \int d^{2} x \epsilon_{ij} \vec{S} \cdot (\partial_{i} \vec{S} \times \partial_{j} \vec{S})
  \end{equation}
  
  is topological charge, $x_{1}=x, \hspace{0.05 in} x_{2}=y$. The resulting Bogomolny equations have the form (derived in the case, when $V=V(S^{3})$), \cite{Adametal2010}
  
  \begin{equation}
 \frac{1}{2} \epsilon_{ij} \vec{S} \cdot ( \partial_{i} \vec{S} \times \partial_{j} \vec{S}) \pm \gamma \sqrt{V(S^{3})} =0. 
 \label{bogomolny_eq_Adam_etal:2010zz}
  \end{equation}

In contrary to \cite{Adametal2010}, we derive Bogomolny equations (we call them as Bogomolny decomposition), by applying so called, concept of strong necessary conditions, firstly presented in \cite{Sokalski1979} and developed in  \cite{Sokalskietal2001}, \cite{Sokalskietal22001}, \cite{Sokalskietal2002}.\\

The procedure of deriving of Bogomolny decomposition from the extended concept of strong necessary conditions, has been presented in \cite{Sokalskietal2002}, \cite{Stepien2003} and developed in \cite{Stepienetal2009}.\\
This paper is organized, as follows. In the next subsections of this section we describe shortly restricted baby Skyrme model, full baby Skyrme model and the concept of strong necessary conditions. In the sections: II and III, we derive Bogomolny decomposition for the baby Skyrme models: restricted and full one, correspondingly, by using the concept of strong necessary conditions. Section IV contains some conclusions.

 \subsection{Baby Skyrme models}

  \begin{enumerate}
  \item restricted baby Skyrme model\\

 The lagrangian of restricted baby Skyrme model follows from the Lagrange density of full baby Skyrme model (\ref{baby_Skyrme}), when the $O(3)$ term is absent, \cite{Adametal2009}, \cite{Adametal2010} 
 
 \begin{equation}
 \mathcal{L} = -\beta(\partial_{\mu} \vec{S} \times \partial_{\nu} \vec{S})^{2} - V(\vec{S}),  \label{lagr_restr}
 \end{equation}

 In this paper we consider the energy functional for restricted baby Skyrme model in (2+0) dimensions, of the following form, 
 \cite{Adametal2010} 
 
 \begin{equation}
 H = \frac{1}{2} \int d^{2} x \mathcal{H} = \frac{1}{2} \int d^{2} x \bigg( \frac{\beta}{4} (\epsilon_{ij} \partial_{i} \vec{S} \times 
 \partial_{j} \vec{S})^{2} + \gamma^{2} V(\vec{S}) \bigg), \label{energy}
 \end{equation}
 
 where $x_{1}=x, \hspace{0.05 in} x_{2}=y$, $\vec{S}$ is three-component vector, such that $\mid \vec{S} \mid^{2} = 1$ and the potential $V$ depends only on $\vec{S}$. We make the stereographic projection
 
 \begin{equation}
 \vec{S} = \bigg[\frac{\omega+\omega^{\ast}}{1+\omega \omega^{\ast}}, \frac{-i(\omega-\omega^{\ast})}{1+\omega \omega^{\ast}}, \frac{1-\omega 
 \omega^{\ast}}{1+\omega \omega^{\ast}}\bigg],   
 \label{stereograf}
 \end{equation} 
 
 where $\omega=\omega(x,y) \in \mathbb{C}$ and $x, y \in \mathbb{R}$.\\ 
 Then, the density of energy functional (\ref{energy}) has the form
 
  \begin{equation}
   \mathcal{H}=-4\beta\frac{(\omega_{,x}\omega^{\ast}_{,y}-\omega_{,y}\omega^{\ast}_{,x})^{2}}{(1+\omega \omega^{\ast})^{4}} + 
   V(\omega,\omega^{\ast}),
  \end{equation}
  
  where $\gamma$ has been included in $V(\omega,\omega^{\ast})$ and $\omega_{,x} \equiv \frac{\partial \omega}{\partial x}$, etc.
  
  The Euler-Lagrange equations for this model are, as follows
 
 \begin{equation}
  \begin{gathered}
  16 \beta\frac{(\omega_{,x}\omega^{\ast}_{,y} - \omega_{,y}\omega^{\ast}_{,x})^{2} \omega^{\ast}}{(1+\omega\omega^{\ast})^{5}} - 
    8 \beta \frac{\omega_{,xx} (\omega^{\ast}_{,y})^{2} + \omega_{,yy} (\omega^{\ast}_{,x})^{2} + (\omega_{,x}\omega^{\ast}_{,y} +  \omega_{,y}\omega^{\ast}_{,x}) \omega^{\ast}_{,xy}}{(1+\omega\omega^{\ast})^{4}} + \\
  8 \beta \frac{2\omega_{,xy}\omega^{\ast}_{,x}\omega^{\ast}_{,y} + \omega_{,x}\omega^{\ast}_{,x}\omega^{\ast}_{,yy} +  \omega_{,y}\omega^{\ast}_{,y}\omega^{\ast}_{,xx}}{(1+\omega\omega^{\ast})^{4}} - V_{,\omega} = 0,\\
  c.c.
   \end{gathered} 
   \end{equation}

 As we mentioned it above, the Bogomolny equations found for this model in (2+0)-dimensions, given by the functional of energy (\ref{energy}), with the potential $V=V(S^{3})$, but by using the technique of proper separation of the terms in the expression for the functional of energy, have the form, \cite{Adametal2010}
 
 \begin{equation}
 \frac{1}{2}\epsilon_{ij}\vec{S}\cdot(\partial_{i}\vec{S} \times \partial_{j}\vec{S}) \pm \gamma \sqrt{V(S^{3})}=0.
 \end{equation}

 Another form of these equations is, as follows, \cite{Adametal2010}
 
 \begin{equation}
 2 \epsilon_{ij}[\partial_{i}(1+\omega\omega^{\ast})^{-1}]\partial_{j} \varphi  \pm 
 \gamma\sqrt{V(\omega\omega^{\ast})}=0, 
 \label{bogomolny2_eq_Adam_etal:2010zz}
 \end{equation} 
 
 where $\varphi=\arg{(\omega)}$. 
 
 It is a generalization of the result obtained in \cite{GP1997}
 
 \begin{equation}
 \frac{1}{2}\epsilon_{ij}\vec{S}\cdot(\partial_{i}\vec{S} \times \partial_{j}\vec{S}) \pm \gamma(\vec{n}-\vec{S})=0
 \end{equation} 
 
 \vspace{0.1 in} 
 
 \item full baby Skyrme model\\
 
 We obtain the full baby Skyrme model, by adding to the lagrangian (\ref{lagr_restr}), the $O(3)$ term: $\partial_{\mu} \vec{S} \cdot \partial^{\mu} \vec{S}$ (here we follow this term by $\alpha$), so we get (\ref{baby_Skyrme}), \cite{Adametal2009}
 
 \begin{equation}
 \mathcal{L}= \alpha \partial_{\mu} \vec{S} \cdot \partial^{\mu} \vec{S} - \beta (\partial^{\mu} \vec{S} \times
 \partial^{\nu} \vec{S})^{2} - \gamma^{2} V(\vec{S}), \label{baby_Skyrme2}
 \end{equation}
 
 where $\alpha, \beta$ are coupling constants.
 
 If we make stereographic projection (\ref{stereograf}), then the density of the functional of energy is, as follows
  
  \begin{equation}
   \mathcal{H} = 4\alpha\frac{\omega_{,x}\omega^{\ast}_{,x}+\omega_{,y}\omega^{\ast}_{,y}}{(1+\omega\omega^{\ast})^{2}} - 4 \beta
   \frac{(\omega_{,x}\omega^{\ast}_{,y}-\omega_{,y}\omega^{\ast}_{,x})^{2}}{(1+\omega\omega^{\ast})^{4}} + \gamma^{2} 
   V(\omega,\omega^{\ast}). \label{full_baby_Skyrme_ster}
  \end{equation}

  It is convenient to write the energy density (\ref{full_baby_Skyrme_ster}) in real field variables $u, v \in 
  \mathbb{R}$: $\omega=u + i v, \omega^{\ast}=u - i v$, include $\gamma^{2}$ in potential $V$ and introduce some constants $\lambda_{1}, 
  \lambda_{2}$
  
   \begin{equation}
   \mathcal{H} = \frac{\lambda_{1}}{2} \frac{u^{2}_{,x} + u^{2}_{,y} + v^{2}_{,x} + 
   v^{2}_{,y}}{(1+u^{2}+v^{2})^{2}} + \lambda_{2} \frac{(u_{,x}v_{,y}-u_{,y}v_{,x})^{2}}{(1+u^{2}+v^{2})^{4}} 
   + V(u,v),
   \end{equation}
   
   where $\lambda_{1}=8\alpha, \lambda_{2}=16\beta$. 
   
  The Euler-Lagrange equations of this model have the following form:
  
  \begin{equation}
  \begin{gathered}
  \lambda_{1}\frac{u_{,xx}+u_{,yy}}{(1+u^{2}+v^{2})^{2}} - 
  2\lambda_{1}\frac{u(u^{2}_{,x}+u^{2}_{,y}-v^{2}_{,x}-v^{2}_{,y})+2v(u_{,x}v_{,x}+u_{,y}v_{,y})}{(1+u^{2}+v^{2})^{3}} +\\ 
  2\lambda_{2}\frac{u_{,xx}v^{2}_{,y}+u_{,yy}v^{2}_{,x}+(u_{,x}v_{,y} + u_{,y}v_{,x})v_{,xy} - 
  2u_{,xy}v_{,x}v_{,y} - u_{,x}v_{,x}v_{,yy} 
  -u_{,y}v_{,y}v_{,xx}}{(1+u^{2}+v^{2})^{4}} -\\ 
  8 \lambda_{2}  \frac{(u_{,x}v_{,y} - u_{,y}v_{,x})^{2} u}{(1+u^{2}+v^{2})^{5}} - V_{,u}=0,\\ 
   \end{gathered} 
   \end{equation}
 
 and the corresponding equation, obtained by varying the functional with respect to $v$.
 
 As it has been stated in \cite{Adametal2010}, the Bogomolny bound for this model cannot be saturated by non-trivial solutions and so, the Bogomolny equations cannot be derived in this case, {\em but by using traditional technique of deriving Bogomolny equations, based on proper separation of the terms in the expression for the functional of energy}. 
 
 \end{enumerate} 
 
 \subsection{The concept of strong necessary conditions}

 The idea of the concept of strong necessary conditions is such that instead of considering of the Euler-Lagrange
 equations, 
 
 \begin{equation}
 F_{,u} - \frac{d}{dx}F_{,u_{,x}} - \frac{d}{dt}F_{,u_{,t}}=0, \label{el}
 \end{equation}
  
 following from the extremum principle, applied to the functional
 
 \begin{equation}
 \Phi[u]=\int_{E^{2}} F(u,u_{,x},u_{,t}) \hspace{0.05 in} dxdt, \label{functional}
 \end{equation}
 
 we consider strong neecessary conditions, \cite{Sokalski1979}, \cite{Sokalskietal2001}, \cite{Sokalskietal22001}, \cite{Sokalskietal2002}
 
 \begin{gather}
   F_{,u}=0, \label{silne1} \\
   F_{,u_{,t}}=0, \label{silne2} \\
   F_{,u_{,x}}=0, \label{silne3}
 \end{gather} 
 
 where $F_{,u} \equiv \frac{\partial F}{\partial u}$, etc.
   
 Obviously, all solutions of the system of the equations (\ref{silne1}) - (\ref{silne3}) satisfy the Euler-Lagrange equation (\ref{el}). However, these solutions, if they exist, are very often trivial. So, in order to avoiding such situation, we make gauge transformation of the functional (\ref{functional})
 
  \begin{equation}
  \Phi \rightarrow \Phi + Inv, \label{gauge_transf}
  \end{equation}
  
 where $Inv$ is such functional that its local variation with respect to $u(x,t)$ vanishes:
 $\delta Inv \equiv 0$. Owing to this feature, the Euler-Lagrange equations (\ref{el}) and the Euler-Lagrange equations resulting from  requiring of the extremum of $\Phi + Inv$, are equivalent. On the other hand, the strong necessary conditions (\ref{silne1}) - (\ref{silne3}) are not invariant with respect to the gauge transformation (\ref{gauge_transf}) and so, we may expect to obtain non-trivial solutions. Let us note that the strong necessary conditions (\ref{silne1}) - (\ref{silne3}) constitute the system of the partial differential equations of the order less than the order of Euler-Lagrange equations (\ref{el}). 

\section{Bogomolny decomposition of restricted baby Skyrme model}

  Now, we apply the concept of strong necessary conditions to the functional (\ref{energy}), in order to find Bogomolny decomposition. We make the following gauge transformation
  
   \begin{equation}
   \mathcal{H} \longrightarrow \tilde{\mathcal{H}}=-4\beta\frac{(\omega_{,x}\omega^{\ast}_{,y}-\omega_{,y}\omega^{\ast}_{,x})^{2}}{(1+\omega 
   \omega^{\ast})^{4}} + V(\omega,\omega^{\ast}) +
   \sum^{3}_{k=1} I_{k}, \label{przecech}
   \end{equation}
  
  where $I_{1}=G_{1}(\omega,\omega^{\ast}) (\omega_{,x}\omega^{\ast}_{,y}-\omega_{,y}\omega^{\ast}_{,x}), I_{2}=D_{x}G_{2}(\omega,\omega^{\ast}), I_{3}=D_{y}G_{3}(\omega,\omega^{\ast}), D_{x} \equiv \frac{d}{dx}, D_{y} \equiv \frac{d}{dy}$ and $G_{k}=G_{k}(\omega,\omega^{\ast}) \in \mathcal{C}^{2}$, ($k=1,2,3$), are some functions, which are to be determinated.
  
  After applying the concept of strong necessary conditions to (\ref{przecech}), we obtain the so-called dual 
  equations
  
  \begin{equation}
  \begin{gathered}
  \tilde{\mathcal{H}}_{,\omega} =
  16 \beta \frac{(\omega_{,x}\omega^{\ast}_{,y}-\omega_{,y}\omega^{\ast}_{,x})^{2}\omega^{\ast}}{(1+\omega 
  \omega^{\ast})^{5}} + V_{,\omega}(\omega,\omega^{\ast}) + \\
  G_{1,w}(\omega,\omega^{\ast})(\omega_{,x}\omega^{\ast}_{,y}-\omega_{,y}\omega^{\ast}_{,x}) + 
  D_{x}G_{2,\omega}(\omega,\omega^{\ast}) + \label{gorne1} \\ 
  D_{y}G_{3,\omega}(\omega,\omega^{\ast})=0,
   \end{gathered} 
   \end{equation}
  
  \begin{equation}
  \begin{gathered}
  \tilde{\mathcal{H}}_{,\omega^{\ast}}=
  16 \beta \frac{(\omega_{,x}\omega^{\ast}_{,y}-\omega_{,y}\omega^{\ast}_{,x})^{2}\omega}{(1+\omega\omega^{\ast})^{5}} 
  + V_{,\omega^{\ast}}(\omega,\omega^{\ast}) + \\ 
  G_{1,\omega^{\ast}}(\omega,\omega^{\ast})(\omega_{,x}\omega^{\ast}_{,y}-\omega_{,y}\omega^{\ast}_{,x}) + 
  D_{x}G_{2,\omega^{\ast}}(\omega,\omega^{\ast}) + 
  \label{gorne2} \\ 
  D_{y}G_{3,\omega^{\ast}}(\omega,\omega^{\ast}) =0, 
  \end{gathered}
  \end{equation}

  \begin{equation}
  \begin{gathered}
  \tilde{\mathcal{H}}_{,\omega_{,x}} = 
  -8 \beta\frac{(\omega_{,x}\omega^{\ast}_{,y}-\omega_{,y}\omega^{\ast}_{,x})\omega^{\ast}_{,y}}{(1+\omega
  \omega^{\ast})^{4}} + G_{1}(\omega,\omega^{\ast})\omega^{\ast}_{,y} +  G_{2,\omega} = 0, \label{dolne1} 
  \end{gathered}
  \end{equation}
  
  \begin{equation}
  \begin{gathered}
  \tilde{\mathcal{H}}_{,\omega_{,y}} 
  = 8 \beta\frac{(\omega_{,x}\omega^{\ast}_{,y}-\omega_{,y}\omega^{\ast}_{,x})\omega^{\ast}_{,x}}{(1+\omega
  \omega^{\ast})^{4}} - G_{1}(\omega,\omega^{\ast})\omega^{\ast}_{,x} +  G_{3,\omega} = 0, \label{dolne2} 
  \end{gathered}
  \end{equation}
  
  \begin{equation}
  \begin{gathered}
  \tilde{\mathcal{H}}_{,\omega^{\ast}_{,x}} 
  = 8 \beta\frac{(\omega_{,x}\omega^{\ast}_{,y}-\omega_{,y}\omega^{\ast}_{,x})\omega_{,y}}{(1+\omega\omega^{\ast})^{4}} 
  - G_{1}(\omega,\omega^{\ast})\omega_{,y} + G_{2,\omega^{\ast}} = 0, \label{dolne3} 
  \end{gathered}
  \end{equation}
  
  \begin{equation}
  \begin{gathered}
  \tilde{\mathcal{H}}_{,\omega^{\ast}_{,y}} = 
   -8 \beta\frac{(\omega_{,x}\omega^{\ast}_{,y}-\omega_{,y}\omega^{\ast}_{,x})\omega_{,x}}{(1+\omega\omega^{\ast})^{4}} + 
  G_{1}(\omega,\omega^{\ast})\omega_{,x} + G_{3,\omega^{\ast}} = 0.  \label{dolne4}
  \end{gathered}
  \end{equation}
  
  \vspace{0.5 in} 
  
  Now, we must make the equations (\ref{gorne1}) - (\ref{dolne4}) self-consistent.
  In this order, we must reduce the number of independent equations by an appropriate choice of the functions $G_{k}, (k =1, 2, 3)$.
  Usually, such ansatzes exist only for some special $V(\omega, \omega^{\ast})$ and in most cases of $V(\omega,\omega^{\ast})$ for many  
  nonlinear field models, it is impossible to reduce the system of corresponding dual equations, to Bogomolny equations. However, even at that time, such system can be used to derive at least some particular set of solutions of Euler-Lagrange equations. \\ 
  
  Now, we consider $\omega, \omega^{\ast}, G_{k}$, ($k=1, 2, 3$), as equivalent dependent variables, governed by the system of equations
  (\ref{gorne1}) - (\ref{dolne4}). We make two operations (they were applied firstly in \cite{Sokalskietal2002} for the cases of  hyperbolic and elliptic systems of nonlinear PDE's). At first, we integrate the equations (\ref{gorne1}) - (\ref{gorne2}) with respect to $\omega$ and to $\omega^{\ast}$, correspondingly. We get
  
  \begin{gather}
  -4\beta \frac{(\omega_{,x}\omega^{\ast}_{,y}-\omega_{,y}\omega^{\ast}_{,x})^{2}}{(1+\omega\omega^{\ast})^{4}} + V(\omega,\omega^{\ast}) + G_{1}(\omega,\omega^{\ast})(\omega_{,x}\omega^{\ast}_{,y}-\omega_{,y}\omega^{\ast}_{,x}) + \nonumber \\ 
 D_{x}G_{2}(\omega,\omega^{\ast}) + D_{y}G_{3}(\omega,\omega^{\ast})=F(\omega_{,x},\omega_{,y},\omega^{\ast}_{,x}, \omega^{\ast}_{,y}),
 \label{po_scalk} 
 \end{gather}
 
 where $F$ is some function, which will be determined later.\\ 
 The second step is making the equations (\ref{dolne1}) - (\ref{dolne4}) self-consistent.
  After proper multiplying of the equations (\ref{dolne1}) - (\ref{dolne4}) by $\omega_{,x}, \omega_{,y}, \omega^{\ast}_{,x}, \omega^{\ast}_{,y}$, correspondingly, and adding by sides the equations (\ref{dolne1}), (\ref{dolne3}) and (\ref{dolne2}), (\ref{dolne4}), we get the relations, including the divergencies $D_{x}G_{2}(w,w^{\ast})$ and $D_{y}G_{3}(w,w^{\ast})$
   
   \begin{gather}
   -8\beta\frac{(\omega_{,x}\omega^{\ast}_{,y}-\omega_{,y}\omega^{\ast}_{,x})^{2}}{(1+\omega\omega^{\ast})^{4}} + 
   G_{1}(\omega,\omega^{\ast})(\omega_{,x}\omega^{\ast}_{,y}-\omega_{,y}\omega^{\ast}_{,x}) + D_{x}G_{2}(\omega,\omega^{\ast}) = 0, 
   \label{dywerg1} \\
   -8\beta\frac{(\omega_{,x}\omega^{\ast}_{,y}-\omega_{,y}\omega^{\ast}_{,x})^{2}}{(1+\omega\omega^{\ast})^{4}} + 
   G_{1}(\omega,\omega^{\ast})(\omega_{,x}\omega^{\ast}_{,y}-\omega_{,y}\omega^{\ast}_{,x}) + D_{y}G_{3}(\omega,\omega^{\ast}) = 0. 
   \label{dywerg2}
   \end{gather}

   Hence
   
   \begin{gather}
   D_{x} G_{2}(\omega,\omega^{\ast})=D_{y}G_{3}(\omega,\omega^{\ast}). \label{z_dywerg}
   \end{gather}

  Moreover, if we multiply again the equations (\ref{dolne1}) - (\ref{dolne4}) by $\omega_{,x}, \omega_{,y}, \omega^{\ast}_{,x}, \omega^{\ast}_{,y}$ and add by sides, but such, that to get the relations, including the divergencies $D_{y}G_{2}(\omega,\omega^{\ast}), D_{x}G_{3}(\omega,\omega^{\ast})$, we get
    
   \begin{gather}
   D_{y}G_{2}(\omega,\omega^{\ast})=0, \hspace{0.08 in} D_{x}G_{3}(\omega,\omega^{\ast})=0. \label{dywerg3}
   \end{gather}
   
   We call the relations (\ref{dywerg1}), (\ref{dywerg2}) and (\ref{dywerg3}), as divergent representation (the  divergent representation was derived firstly in \cite{Sokalskietal2002} for hyperbolic system of two coupled nonlinear partial differential equations). 
   
   Hence, and from (\ref{z_dywerg})
   
   \begin{gather}
   G_{2}(\omega,\omega^{\ast})=const, \hspace{0.1 in} G_{3}(\omega,\omega^{\ast})=const. \label{G2_G3}
   \end{gather}

  Then, from the relation (\ref{dywerg1}) (or (\ref{dywerg2})), we have
  
  \begin{gather}
   -8\beta\frac{(\omega_{,x}\omega^{\ast}_{,y}-\omega_{,y}\omega^{\ast}_{,x})^{2}}{(1+\omega\omega^{\ast})^{4}} +  
   G_{1}(\omega,\omega^{\ast})(\omega_{,x}\omega^{\ast}_{,y}-\omega_{,y}\omega^{\ast}_{,x}) = 0, \label{r34}
   \end{gather}
  
  From (\ref{r34}) we get
  
  \begin{gather}
  \omega_{,x}\omega^{\ast}_{,y}-\omega_{,y}\omega^{\ast}_{,x} = \frac{1}{8\beta} G_{1}(\omega,\omega^{\ast})(1+\omega\omega^{\ast})^{4}. 
  \label{bogomolny_G}
  \end{gather}
  
  We obtain the same result from (\ref{dolne1})-(\ref{dolne4}).
  One can easily check that all solutions of (\ref{bogomolny_G}) satisfy the equations (\ref{dolne1}) - (\ref{dolne4}).
  Now, we must investigate, when the equation (\ref{po_scalk}) is satisfied by the solutions of (\ref{bogomolny_G}). 
  Then, we insert (\ref{G2_G3}) and (\ref{bogomolny_G}), into the equation (\ref{po_scalk})
  
  \begin{gather}
  V(\omega,\omega^{\ast})+\frac{1}{16 \beta}G_{1}^{2}(\omega,\omega^{\ast})(1+\omega\omega^{\ast})^{4}=F(\omega_{,x},\omega_{,y},
  \omega^{\ast}_{,x}, \omega^{\ast}_{,y}). \label{pot1}
  \end{gather}
  
  Now, in order to determining function $F$, we compare (\ref{pot1}) with Hamilton-Jacobi equation, which has the form, 
  \cite{Sokalskietal2002}
  
  \begin{equation}
  \tilde{\mathcal{H}}=0, \label{HJ} 
  \end{equation}
  
  where, of course $\tilde{\mathcal{H}}$ in general, for $\omega=\omega(x^{\mu}), \omega^{\ast}=\omega^{\ast}(x^{\mu})$, ($\mu = 0,1,2,3$ and 
  $x^{0}=t$), is defined, as follows
  
  \begin{equation}
  \tilde{\mathcal{H}}=\Pi_{\omega} \omega_{,t} + \Pi_{\omega^{\ast}} \omega^{\ast}_{,t} - \tilde{\mathcal{L}}
  \end{equation}

  and $\Pi_{\omega}=\tilde{\mathcal{L}}_{\omega_{,t}}, \Pi_{\omega^{\ast}}=\tilde{\mathcal{L}}_{\omega^{\ast}_{,t}}$ are canonical 
  momenta and $\tilde{\mathcal{L}}$ is Lagrange density gauge-transformed on the invariants $I_{k}, (k=1, 2, 3)$.\\ 
  
  In our case
  
   \begin{equation}
  \tilde{\mathcal{H}}= - \tilde{\mathcal{L}}. \label{lagr_hamilt} 
  \end{equation}

  By inserting into this equation, the relations (\ref{G2_G3}) and (\ref{bogomolny_G}), and taking into account (\ref{HJ}), we get that $F=0$.
  Hence, we get

  \begin{gather}
  V(\omega,\omega^{\ast})=-\frac{1}{16\beta} G_{1}^{2}(\omega,\omega^{\ast})(1+\omega\omega^{\ast})^{4}. \label{warunek_pot}
  \end{gather}
  
  Then, of course, 
  
  \begin{equation}
  G_{1}=\frac{4i\sqrt{\beta}}{(1+\omega\omega^{\ast})^{2}}\sqrt{V(\omega,\omega^{\ast})}. \label{warunek_G}
  \end{equation} 
  
  We insert (\ref{warunek_G}) in (\ref{bogomolny_G}) and we obtain Bogomolny decomposition for the given potential $V(w,w^{\ast})$
  
  \begin{gather}
  \omega_{,x}\omega^{\ast}_{,y}-\omega_{,y}\omega^{\ast}_{,x} = \frac{i}{2\sqrt{\beta}} \sqrt{V(\omega, \omega^{\ast})} 
  (1+\omega\omega^{\ast})^{2}. 
  \label{bogomolny_decomp}
   \end{gather}
  
  Then, the equation (\ref{bogomolny_decomp}) is Bogomolny decomposition (Bogomolny equation) for restricted baby Skyrme model in (2+0) 
  dimensions, for {\em arbitrary} potential.\\
  
  \vspace{2 in}

  \section{The search for Bogomolny decomposition of the full baby Skyrme model}
  
  Now we apply the concept of strong necessary conditions to the functional (\ref{full_baby_Skyrme_ster}), in order to find Bogomolny 
  decomposition for full baby Skyrme model in (2+0)-dimensions. We make gauge transformation, but as follows
  
   \begin{equation}
   \begin{gathered}
   \mathcal{H} \longrightarrow \tilde{\mathcal{H}}= \frac{\lambda_{1}}{2} \frac{u^{2}_{,x} + u^{2}_{,y} + 
   v^{2}_{,x} + v^{2}_{,y}}{(1+u^{2}+v^{2})^{2}} + \lambda_{2} \frac{(u_{,x}v_{,y}-u_{,y}v_{,x})^{2}}{(1+u^{2}+v^{2})^{4}} + \\
    V(u,v) +  G_{1}(u,v) (u_{,x}v_{,y}-u_{,y}v_{,x}) + \label{full_przecech} \\ 
    H_{1}(u,v) (u_{,x}v_{,y}-u_{,y}v_{,x}) + D_{x}G_{2}(u,v) + \\
    D_{y}G_{3}(u,v) + D_{x}H_{2}(u,v) + D_{y}H_{3}(u,v), 
   \end{gathered}
   \end{equation}
  
  where $G_{k}(u,v), H_{k}(u,v) \in \mathcal{C}^{2} $, ($k=1,2,3$), are some functions, which are to be determined later.
  
  After applying the concept of strong necessary conditions to (\ref{full_przecech}), we obtain the dual 
  equations
  
  \begin{equation}
  \begin{gathered}
  \tilde{\mathcal{H}}_{,u} = - 2 \lambda_{1} \frac{(u^{2}_{,x} + u^{2}_{,y} + v^{2}_{,x} + 
  v^{2}_{,y})u}{(1+u^{2}+v^{2})^{3}} - 8 \lambda_{2} \frac{(u_{,x}v_{,y}-u_{,y}v_{,x})^{2}u}{(1+u^{2}+v^{2})^{5}} + \\  
  V_{,u} + G_{1,u} (u_{,x}v_{,y}-u_{,y}v_{,x}) + H_{1,u} (u_{,x}v_{,y}-u_{,y}v_{,x}) + \label{full_gorne1} \\
  D_{x}G_{2,u} + D_{y}G_{3,u} + D_{x}H_{2,u} + D_{y}H_{3,u} =0, 
  \end{gathered} 
  \end{equation}
  
  \begin{equation}
  \begin{gathered}
  \tilde{\mathcal{H}}_{,v} = - 2 \lambda_{1} \frac{(u^{2}_{,x} + u^{2}_{,y} + v^{2}_{,x} + 
  v^{2}_{,y})v}{(1+u^{2}+v^{2})^{3}} - 8\lambda_{2}\frac{(u_{,x}v_{,y}-u_{,y}v_{,x})^{2}v}{(1+u^{2}+v^{2})^{5}} + \\
  V_{,v} +  G_{1,v} (u_{,x}v_{,y}-u_{,y}v_{,x}) + H_{1,v} (u_{,x}v_{,y}-u_{,y}v_{,x}) +  \\
  D_{x}G_{2,v} + D_{y}G_{3,v} + D_{x}H_{2,v} + D_{y}H_{3,v}=0, \label{full_gorne2}  
  \end{gathered}
  \end{equation}
  
  \begin{equation}
  \begin{gathered}
  \tilde{\mathcal{H}}_{,u_{,x}} = \lambda_{1} \frac{u_{,x}}{(1+u^{2}+v^{2})^{2}} + 
  2\lambda_{2}\frac{(u_{,x}v_{,y}-u_{,y}v_{,x})v_{,y}}{(1+u^{2}+v^{2})^{4}} + \\ 
  G_{1}v_{,y} + H_{1} v_{,y} + G_{2,u} + H_{2,u} = 0, \label{full_dolne1}   
   \end{gathered}
  \end{equation}
  
  \begin{equation}
  \begin{gathered}
  \tilde{\mathcal{H}}_{,u_{,y}} = \lambda_{1}\frac{u_{,y}}{(1+u^{2}+v^{2})^{2}} - 
  2\lambda_{2}\frac{(u_{,x}v_{,y}-u_{,y}v_{,x})v_{,x}}{(1+u^{2}+v^{2})^{4}} - \\
  G_{1} v_{,x} - H_{1} v_{,x} +  G_{3,u} + H_{3,u} = 0, \label{full_dolne2} \\ 
  \end{gathered}
  \end{equation}
  
  \begin{equation}
  \begin{gathered}
  \tilde{\mathcal{H}}_{,v_{,x}} = \lambda_{1} \frac{v_{,x}}{(1+u^{2}+v^{2})^{2}} - 2 \lambda_{2} 
  \frac{(u_{,x}v_{,y}-u_{,y}v_{,x})u_{,y}}{(1+u^{2}+v^{2})^{4}} - \\
  G_{1} u_{,y} - H_{1} u_{,y} + G_{2,v} + H_{2,v} = 0, \label{full_dolne3} \\ 
   \end{gathered}
  \end{equation}
  
  \begin{equation}
  \begin{gathered}
  \tilde{\mathcal{H}}_{,v_{,y}} = \lambda_{1} \frac{v_{,y}}{(1+u^{2}+v^{2})^{2}} + 2 \lambda_{2} 
  \frac{(u_{,x}v_{,y}-u_{,y}v_{,x})u_{,x}}{(1+u^{2}+v^{2})^{4}} + \\
   G_{1} u_{,x} + H_{1} u_{,x} + G_{3,v} + H_{3,v} = 0, \label{full_dolne4} \\
  \end{gathered}
  \end{equation}
  
  where, of course, $G_{k}, H_{k} \in \mathcal{C}^{2}, (k=1,2,3)$, are some functions of $u,v$, mentioned above. 
  
  Firstly, we integrate (\ref{full_gorne1})-(\ref{full_gorne2}) with respect to $u$ and $v$, 
  correspondingly
  
  \begin{equation}
  \begin{gathered}
  \frac{\lambda_{1}}{2} \frac{(u^{2}_{,x}+u^{2}_{,y}+v^{2}_{,x}+v^{2}_{,y})}{(1+u^{2}+u^{2})^{2}}
  + \lambda_{2} \frac{(u_{,x}v_{,y}-u_{,y}v_{,x})^{2}}{(1+u^{2}+u^{2})^{4}} + \\
  G_{1}(u,v)(u_{,x}v_{,y}-u_{,y}v_{,x}) + D_{x}G_{2}(u,v) + \label{full_po_scalk} \\ 
  D_{y}G_{3}(u,v) + H_{1}(u,v)(u_{,x}v_{,y}-u_{,y}v_{,x}) + D_{x} H_{2}(u,v) + \\
  D_{y} H_{3}(u,v) + V(u,v)=F(u_{x,}, u_{,y}, v_{,x}, v_{,y}),
  \end{gathered}
  \end{equation}

  where $F$ is some function of $u_{x,}, u_{,y}, v_{,x}, v_{,y}$. Now, the first step of making the equations 
  (\ref{full_dolne1})-(\ref{full_dolne4}) consistent, is putting
  
  \begin{gather}
  u_{,x}v_{,y}-u_{,y}v_{,x} = -\frac{1}{2\lambda_{2}}(1+u^{2}+v^{2})^{4} G_{1}(u,v) , \label{Bogomolny_eq1} \\
  G_{2}(u,v) = const., \hspace{0.1 in} G_{3}(u,v)=const. \label{war_G1G2}
  \end{gather} 
  
  So, we have here the Bogomolny equation for restricted baby Skyrme model, as the first of wanted Bogomolny 
  equations for full baby Skyrme model.
  However, it is not all, we need to make the second step: choosing the functions $H_{1}, H_{2}, H_{3}$, such that the equations 
  (\ref{full_dolne1})-(\ref{full_dolne4}) will be consistent, when (\ref{Bogomolny_eq1}) - (\ref{war_G1G2}) are satisfied. 
  
 In \cite{Stepien2003} and in \cite{Stepienetal2009} , the Bogomolny decompositions for the model generating generalized parabolic systems of NPDE's of second order, were found (in \cite{Stepienetal2009} Bogomolny decomposition was found by applying not divergent representation, but in other way). We apply here some results from \cite{Stepien2003}, in order to make the equations (\ref{full_gorne1})-(\ref{full_dolne4}) consistent, because the hamiltonian (\ref{full_baby_Skyrme_ster}), but with $\lambda_{2}=0$, corresponds to the hamiltonian of Heisenberg model of ferromagnet in (2+0)-dimensions and it is some special case of the model investigated in \cite{Stepien2003} and \cite{Stepienetal2009}. If (\ref{Bogomolny_eq1})-(\ref{war_G1G2}) are satisfied, then the system (\ref{full_dolne1})-(\ref{full_dolne4}) is special case of the corresponding system of equations, for which, in \cite{Stepien2003}, the divergent representation has been derived. Then, if we apply the results from \cite{Stepien2003}, then the divergent representation for (\ref{full_dolne1})-(\ref{full_dolne4}) (when (\ref{Bogomolny_eq1}) - (\ref{war_G1G2}) are satisfied) has the form
  
  \begin{gather}
  \lambda_{1} \frac{u^{2}_{,x} + v^{2}_{,x}}{(1+u^{2}+v^{2})^{2}} +  
  H_{1}(u_{,x}v_{,y}-u_{,y}v_{,x}) = -D_{x} H_{2}, \label{full_dywerg1} \\
  \lambda_{1} \frac{u_{,x}u_{,y} + v_{,x}v_{,y}}{(1+u^{2}+v^{2})^{2}}  = -D_{y} H_{2}, \label{full_dywerg2} \\
  \lambda_{1} \frac{u_{,x}u_{,y} + v_{,x}v_{,y}}{(1+u^{2}+v^{2})^{2}}  = -D_{x} H_{3}, \label{full_dywerg3} \\
  \lambda_{1} \frac{u^{2}_{,y} + v^{2}_{,y}}{(1+u^{2}+v^{2})^{2}} + H_{1}(u_{,x}v_{,y}-u_{,y}v_{,x}) = -D_{y} H_{3}.  
  \label{full_dywerg4}
  \end{gather}
  
  \vspace{0.5 in} 
  
  From (\ref{full_dywerg2})-(\ref{full_dywerg3}), we get
  
  \begin{equation}
  D_{y} H_{2} = D_{x} H_{3}.
  \end{equation}
  
  \vspace{0.1 in} 
  
  If we put 
  
  \begin{equation}
  H_{1}=\frac{\lambda_{1}}{(1+u^{2}+v^{2})^{2}} \label{war_H1}
  \end{equation} 
  
  and 
  
  \begin{equation}
   H_{2,u} = H_{3,v}, \hspace{0.05 in} H_{2,v} = -H_{3,u}, \label{pierw_poch_H2}
  \end{equation}
  
  and we take into account (\ref{Bogomolny_eq1})-(\ref{war_G1G2}), then, the equations (\ref{full_dolne1})-(\ref{full_dolne4}) will be 
  reduced to
  
  \begin{gather}
  \lambda_{1} \frac{u_{,x}+v_{,y}}{(1+u^{2}+v^{2})^{2}} = -H_{2,u}, \label{Bogomolny_eq2} \\
  \lambda_{1} \frac{u_{,y}-v_{,x}}{(1+u^{2}+v^{2})^{2}} = H_{2,v}. \label{Bogomolny_eq3}
  \end{gather} 
  
  From (\ref{pierw_poch_H2}) it follows that $H_{2}(u,v)$ must satisfy Laplace equation
  
  \begin{equation}
  H_{2,uu} + H_{2,vv} = 0. \label{rown_Laplace}
  \end{equation}
  
  Now, in order to determine $F$ in (\ref{full_po_scalk}), we use Hamilton-Jacobi equation
  
  \begin{equation}
  \tilde{\mathcal{H}}=0, \label{HJ2} 
  \end{equation}
  
  where, of course, $\tilde{\mathcal{H}}$ in general, for $u=u(x^{\mu}), v=v(x^{\mu})$, ($\mu = 0,1,2,3$), 
  is defined, as
  
  \begin{equation}
  \tilde{\mathcal{H}}=\Pi_{u} u_{,t} + \Pi_{v} v_{,t} - \tilde{\mathcal{L}}
  \end{equation}

  and $\Pi_{u}=\tilde{\mathcal{L}}_{u_{,t}}, \Pi_{v}=\tilde{\mathcal{L}}_{v_{,t}}$ are canonical 
  momenta and $\tilde{\mathcal{L}}$ is Lagrange density gauge-transformed in the same way, as the hamiltonian
  (\ref{full_przecech}). In our case
  
   \begin{equation}
  \tilde{\mathcal{H}}= - \tilde{\mathcal{L}}. 
  \end{equation}
  
  Thus, we get that $F(u_{x,}, u_{,y}, v_{,x}, v_{,y})=0$. 
  After taking into consideration (\ref{Bogomolny_eq1}) - (\ref{war_G1G2}), (\ref{war_H1}) and (\ref{Bogomolny_eq2})-(\ref{Bogomolny_eq3}), 
  we obtain the condition for the potential $V(u,v)$ 
  
    \begin{equation}
    \begin{gathered}
     V(u,v)=\frac{(1+u^{2}+v^{2})^{4}}{4\lambda_{2}} G^{2}_{1}(u,v) + \frac{(1+u^{2}+v^{2})^{2}}{2\lambda_{1}}\bigg[H^{2}_{2,u}(u,v) + 
     H^{2}_{2,v}(u,v)\bigg], \label{potencjal_full}
    \end{gathered} 
    \end{equation} 
    
    \vspace{0.2 in}
    
    where $G_{1}(u,v) \in \mathcal{C}^{2}$ and $H_{2}(u,v)$ is some solution of Laplace equation (\ref{rown_Laplace}).\\
    
    Then, we have obtained some system of three equations of first-order
    
    \begin{gather}
    u_{,x}v_{,y}-u_{,y}v_{,x} = -\frac{1}{2\lambda_{2}} (1+u^{2}+v^{2})^{4} G_{1}(u,v), \label{Bogomolny_full1} \\
    u_{,x}+v_{,y} = -\frac{1}{\lambda_{1}}(1+u^{2}+v^{2})^{2} H_{2,u}(u,v), \label{Bogomolny_full2} \\
    u_{,y}-v_{,x} = \frac{1}{\lambda_{1}}(1+u^{2}+v^{2})^{2} H_{2,v}(u,v), \label{Bogomolny_full3}
    \end{gather} 
    
    which includes two uknown functions $u, v$ of two independent variables $x,y$. These equations (\ref{Bogomolny_full1})-(\ref{Bogomolny_full3}) constitute, with the conditions (\ref{rown_Laplace}) and (\ref{potencjal_full}), so called Bogomolny relationship (this notion was used firstly in \cite{Stepien2003}), for full baby Skyrme model in (2+0)-dimensions. These same results can be obtained, by applying some results from \cite{Stepienetal2009}.
 
    \vspace{0.2 in} 
   
   \section{Summary}
   
   We derived Bogomolny decomposition for both: restricted baby Skyrme model and full baby Skyrme model, in (2+0)-dimensions, by using the concept of strong necessary conditions. We see that Bogomolny equation (\ref{bogomolny_decomp}) for this first model of mentioned ones above, is a generalization of the Bogomolny equation (\ref{bogomolny2_eq_Adam_etal:2010zz}), obtained in \cite{Adametal2010}. 
   In \cite{Speight2010} Bogomolny bound (and Bogomolny equation resulting from it) for the energy of restricted baby Skyrme model for the potential of the form $V=\frac{1}{2}U^{2}$, (where $U$ is a non-negative function of class $\mathcal{C}^{1}$ on $N=S^{2} \subset \mathbb{R}^{3}$ with isolated zeroes), was obtained, in the language of differential forms, not by using the concept of strong necessary conditions. We stress here that the Bogomolny decomposition (\ref{bogomolny_decomp}) has been obtained {\em without any assumption} of the form of the potential.\\
    The Bogomolny decomposition, for full baby Skyrme model in  (2+0)-dimensions, consists of three PDE's of first order, for two unknown functions of two independent variables. If we look on the system (\ref{Bogomolny_full1})-(\ref{Bogomolny_full3}), then the first equation of this system is the Bogomolny decomposition of restricted baby Skyrme model. Next, in order to make the system of dual equations self-consistent, we must add to this equation (\ref{Bogomolny_full1}), two equations (\ref{Bogomolny_full2}) and (\ref{Bogomolny_full3}), containing the derivatives of $H_{2}$. Hence, the equation (\ref{Bogomolny_full1}) is the Bogomolny-decomposition for restricted baby Skyrme model, which is constrained by additional equations (\ref{Bogomolny_full2})-(\ref{Bogomolny_full3}) and by the conditions: (\ref{rown_Laplace}), (\ref{potencjal_full}). Thus, the set of solutions of the Bogomolny decomposition for full baby Skyrme model is a subset of the set of solutions of the Bogomolny decomposition for  restricted baby Skyrme model. Moreover, as far as we derived Bogomolny decomposition for restricted baby Skyrme model for {\em arbitrary} potential $V(u,v)$, the Bogomolny decomposition for full baby Skyrme model can be derived only for the class of potentials, given by (\ref{potencjal_full}), when (\ref{rown_Laplace}) is satisfied.\\ 
   
   So, by adding the additional term in lagrangian of restricted baby Skyrme model, in order to get full baby Skyrme model, we cannot get more wide set of solutions of Bogomolny decomposition. It corresponds to the fact that non-trivial solutions of full baby Skyrme model cannot saturate Bogomolny bound, which was noticed in \cite{Adametal2010}, because this fact concerned the trials of derivation of Bogomolny bound for full baby Skyrme model, which had been made for {\em  arbitrary} potential. The technique of deriving of Bogomolny equations, used for baby Skyrme models in \cite{Adametal2010}, is based on the idea, applied firstly by Bogomolny in \cite{Bogomolny1976}, which gave many important results and belongs to classical methods in nonlinear field theory. However, there exist some classes of field models, (for e.g. full baby Skyrme model) for which Bogomolny equations cannot be derived by this method and we must apply here the concept of strong necessary conditions.

  \section{Computational resources} 
  
   The computations were carried out, by using Waterloo MAPLE 12 Software on the computer "mars"
   (No. of grants: MNiI/IBM BC HS21/AP/057/2008). The computations were carried out also in Interdisciplinary 
   Centre for Mathematical and Computer Modelling (ICM), within the grant No. G31-6.

 \section{Acknowledgments}
  The author thanks to Dr A. Wereszczy\'{n}ski for discussion.



 \bibliographystyle{unsrt}

\bibliography{LTStepien}

\end{document}